\documentclass[%
 aip,
 amsmath,amssymb,
 reprint,%
]{revtex4-1}

\usepackage{graphicx}
\usepackage{dcolumn}
\usepackage{bm}

\usepackage[utf8]{inputenc}
\usepackage[T1]{fontenc}
\usepackage{mathptmx}
\usepackage{etoolbox}

\usepackage{xcolor}
\usepackage{amssymb}
\usepackage{comment}
\usepackage{hyperref}
\usepackage{wasysym}
\DeclareUnicodeCharacter{03DE}{\lightning}

\makeatletter
\def\@email#1#2{%
 \endgroup
 \patchcmd{\titleblock@produce}
  {\frontmatter@RRAPformat}
  {\frontmatter@RRAPformat{\produce@RRAP{*#1\href{mailto:#2}{#2}}}\frontmatter@RRAPformat}
  {}{}
}%
\makeatother
\begin{document}

\preprint{AIP/123-QED}

\title{Recovery Random Walks for Extreme Events on Complex Networks}
\author{Karan Singh}
 \affiliation{School of Physics, Indian Institute of Science Education and Research Thiruvananthapuram, Thiruvananthapuram, 695551, Kerala, India.}
 
\author{Narendran R. V.}
 \affiliation{School of Physics, Indian Institute of Science Education and Research Thiruvananthapuram, Thiruvananthapuram, 695551, Kerala, India.}
 
\author{V. K. Chandrasekar}
 \affiliation{Department of Physics, Center for Nonlinear Science and Engineering, School of Electrical and Electronics Engineering, SASTRA Deemed University, Thanjavur, 613401, Tamil Nadu, India.}

\author{D. V. Senthilkumar}
 \email{skumar@iisertvm.ac.in}
 \affiliation{School of Physics, Indian Institute of Science Education and Research Thiruvananthapuram, Thiruvananthapuram, 695551, Kerala, India.}

\begin{abstract}
Extreme events are widely studied within simple random walk frameworks, where their probability is determined by the network structure and stationary walker distribution. Here, we propose a recovery random walk (RRW) model in which extreme events temporally `freeze' the nodes where they occur for a fixed duration $\Delta$ ($\Delta=0$ recovers the original model), trapping walkers and reducing the effective mobile population, thereby making the model more practical. We derive a first-principles description of this feedback and a delayed differential equation for the frozen-node fraction. This finite freezing produces an initial overshoot, followed by damped oscillatory relaxation to a steady state for the fraction of the frozen nodes. We find that freezing suppresses extreme event probability while retaining its degree dependence dynamics, and suppresses the EE frequency. Further, the analytical prediction for the fraction of the frozen nodes agrees closely with simulations. This model closely reflects real-world scenarios, yielding more practical EE statistics.
\end{abstract}

\maketitle

\begin{quotation}
Random walks on complex networks provide a useful tool for studying the occurrence of extreme events (EEs), such as unusually high concentrations of vehicles at road intersections or data packets at communication nodes. In conventional random walk models, however, these accumulations do not alter the dynamics of the walkers. In real-world systems, excessive accumulation can restrict movement, disrupt traffic flow, and influence subsequent walker dynamics, thereby introducing feedback between extreme events and the processes that generate them. To capture this feedback, we introduce a recovery random walk (RRW) model in which nodes experiencing extreme events are temporarily frozen, obstructing walker movement through the network. After a prescribed duration, these nodes recover, with trapped walkers gradually released during the recovery process. This mechanism allows extreme events to influence subsequent network activity rather than remaining passive fluctuations in walker occupation. We investigate how the interplay between extreme events, congestion, and recovery reshapes the temporal dynamics of extreme events on complex networks.
\end{quotation}

\section{Introduction}
\label{sec:Intro}

Random walks provide one of the simplest stochastic descriptions of transport on complex networks, in which an entity moves between nodes according to prescribed transition probabilities. Random walks form the basis for a wide range of network processes and have been used to study diffusion~\cite{masuda2017,newman2003}, first-passage processes~\cite{lovasz1993,tejedor2009}, information flow~\cite{boccaletti2006}, centrality~\cite{nohrieger2004}, community structure~\cite{ponslatapy2006,rosvall2008}, ranking~\cite{brinpage1998,kleinberg1999}, and graph neural networks~\cite{nikolentzos2020random,jin2022rawgnn,naren2026pdrwgcn}. Their generality makes them a natural framework in studying transport on heterogeneous networks, where the underlying topology can strongly influence the resulting dynamics~\cite{boccaletti2006,nicolaides2010,brockmann2006,song2010,barthelemy2004}.
\\

Extreme Events (EE) are rare events in which the occupancy of a node exceeds a prescribed threshold, and such events can recur over time across a wide range of natural, technological, and social systems~\cite{albeverio2005,eichner2007,santhanam2008}. In the context of network transport, such events can represent phenomena like traffic congestion~\cite{demartino2009,echenique2005} and excessive loads on communication systems~\cite{germano2006,wang2009}. In a seminal work, Kishore \textit{et al.} introduced a random-walk framework for studying EEs on complex networks and showed the counterintuitive result that low-degree nodes have a higher probability of experiencing EEs than highly-connected hubs~\cite{kishore2011}. Subsequent studies have extended this framework to degree-biased walks~\cite{kishore2012}, event-size fluctuations~\cite{kishore2012,meloni2008}, capacity manipulation~\cite{argollo2004}, and other transport settings~\cite{eichner2007,santhanam2008}. These studies have established how network topology, transport rules and prescribed thresholds determine the probability and recurrence of EEs, but the occurrence of an EE itself does not, in general, change the subsequent transport dynamics in the system. Particularly, the existing random walk EE framework treats the walker population as continually available for transport regardless of an EE. This differs from many physical transport processes in which an extreme local load can restrict mobility temporarily~\cite{echenique2005,kim2009,tadic2004,demoura2005}, reduce effective capacity~\cite{demartino2009,wang2009}, or induce congestion and jamming~\cite{danila2006,zhao2005}. Related studies of congestion and jamming have incorporated finite transport capacity~\cite{demartino2009,echenique2005,kim2009,demoura2005} and collective traffic limitations~\cite{tadic2007,germano2006,wang2009,danila2006}, but a direct coupling between an EE and a finite-time reduction in the mobile walker population remains largely unexplored.
\\

In this work, we introduce the recovery random walk (RRW) model, in which an EE temporarily `freezes' the nodes at which it occurs and immobilizes the walkers present there for a prescribed duration ($\Delta$). The resulting model provides a more realistic description of transport systems in which local accumulation of walkers due to EEs can temporarily reduce accessibility or mobility. We find that increasing the freezing duration suppresses the overall EE probability while preserving its underlying degree dependence dynamics inherited from the underlying simple random walk, and that the finite recovery time produces delayed relaxation and transient oscillations in the fraction of the frozen nodes. This freezing process thus creates a feedback between EE generation and subsequent transport activity, while the finite recovery time introduces memory into the coarse-grained dynamics. We derive a delayed differential equation for the frozen-node population and obtain a first-principles description that closely reproduces the observed dynamics. 
\\

The remainder of the paper is organized as follows. Section~\ref{sec:Model} starts with a recap of the baseline model with unbiased random walk and definition of EEs from Ref.\cite{kishore2011}, then introduces our RRW model, the freezing mechanism, sequential release of trapped walkers, and the resulting degree-dependent statistics. Section~\ref{sec:Analytic} develops the analytical framework, including the effective mobile walker population, the network-averaged freezing rate, and the delayed evolution equation for the fraction of frozen nodes. Section~\ref{sec:Results} validates these predictions against numerical simulations and examines the degree-dependent, temporal, and microscopic organization of EEs. Finally, Section~\ref{sec:DiscConc} discusses the physical significance, limitations and possible extensions of the RRW model, and summarizes the main conclusions.

\section{Model}
\label{sec:Model}

\subsection{Unbiased Random Walks and Extreme Events}
We consider a connected, undirected network $G(N,E)$ consisting of $N$ nodes and $E$ edges. The degree of node $i$ is denoted by $K_i=\sum_j A_{ij}$, where $A$ is the adjacency matrix of the network. A walker located at node $i$ moves to one of its neighboring nodes with equal probability. Thus, the transition probability from node $i$ to node $j$ is 
\begin{equation}
T_{ij} = P(i\rightarrow j) = \frac{A_{ij}}{K_i}.
\label{eq:transition}
\end{equation}

The resulting stationary occupation probability is~\cite{nohrieger2004}
\begin{equation}
p_i = \frac{K_i}{2E}.
\label{eq:stationary}
\end{equation}
Thus, the stationary occupation probability of a node is determined solely by its degree.

For $W_0$ statistically independent walkers, the instantaneous number of walkers $f_i$ at node $i$ follows a binomial distribution 
with mean and variance~\cite{kishore2011}
\begin{equation}
\langle f_i\rangle = W_0p_i \quad \text{and} \quad \sigma_i^2 = W_0p_i(1-p_i),
\label{eq:mean_occupation_and_variance}
\end{equation}
where $p_i$ is taken from Eq.~\eqref{eq:stationary}.

Following the extreme-event framework of Kishore \textit{et al.}~\cite{kishore2011}, an extreme event is defined as an instantaneous node occupation that exceeds a prescribed threshold,
\begin{equation}
f_i(t)>q_i
\label{eq:ee_condition}
\end{equation}
the threshold is defined as
\begin{equation}
q_i = W_0p_i + M\sqrt{W_0p_i(1-p_i)},
\label{eq:threshold}
\end{equation}
where $M$ controls the rarity of the event. In all simulations, we use $M=4$~\cite{kishore2011}. The threshold $q_i$ is determined from the initial walker population $W_0$ and remains fixed throughout the dynamics.

For the baseline case, $\Delta=0$, the probability of an extreme event at a node depends only on its degree through the stationary occupation probability. Following Kishore \textit{et al.}~\cite{kishore2011}, the degree-dependent EE probability can be written as
\begin{equation}
F(K) = \sum_{f=\lfloor q\rfloor+1}^{W_0} \binom{W_0}{f} p^f(1-p)^{W_0-f},
\label{eq:FK_unbiased}
\end{equation}
where $p$ and $q$ are taken from Eq.~\eqref{eq:stationary} and Eq.~\eqref{eq:threshold}, respectively, which provides the analytical framework for the EE probability under an unbiased random walk.

\subsection{Recovery Random-Walk Dynamics}
We now  modify  the standard random walk model where the occurrence of an EE affects the subsequent walker dynamics. In contrast to the conventional random walk framework, where walkers remain mobile throughout the simulation, here an EE occurring at a node temporally `freezes' the node for a finite time duration $\Delta$, for which the walkers on the frozen node do not participate in the random-walk dynamics. After the freezing period, the node is `unfrozen' and the walkers resume their motion according to the transition probability in Eq.~\eqref{eq:transition} resulting in the proposed recovery random walk (RRW) model. The limit $\Delta=0$ recovers the original random-walk EE model~\cite{kishore2011}.


\subsection{EE Detection and Freezing}
EEs are evaluated only on active (i.e. unfrozen) nodes, meaning a node can generate a new EE only when its occupation exceeds its threshold and its freeze timer has expired. Denoting the remaining freeze time of node $i$ by $\tau_i(t)$, a new EE is registered as
\begin{equation}
\mathrm{EE}_i(t) =
\mathbb{I}\!\left[f_i(t)>q_i\right]
\mathbb{I}\!\left[\tau_i(t)=0\right],
\label{eq:new_ee_condition}
\end{equation}
where $\mathbb{I}[\cdot]$ denotes the indicator function.

When $\mathrm{EE}_i(t)=1$, node $i$ registers an EE, which is then frozen for a  subsequent $\Delta$ time steps. Frozen nodes cannot generate subsequent EEs until released, and hence repeated threshold exceedance during the frozen period are not counted as new EEs.

\begin{figure*}[t]
    \centering
    \includegraphics[width=\textwidth]{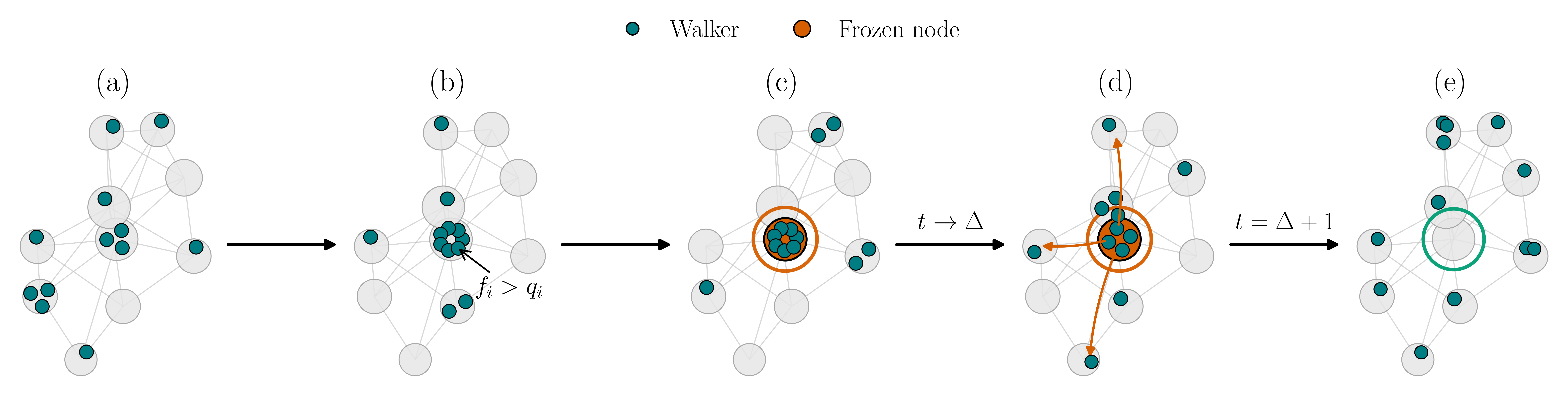}
    \caption{Schematic of the RRW dynamics with EE-induced freezing: (a) Walkers undergo normal random walk dynamics on the BA network; (b) An EE occurs when the flux through node $i$ exceeds its threshold $(f_i>q_i)$; (c) The node then freezes and the walkers at that node are held for a duration $\Delta$; (d) As the freezing interval approaches its end, the held walkers are released sequentially; (e) At $t=\Delta+1$, the node unfreezes (shown enclosed by a green ring) and normal random-walk dynamics resumes.}
    \label{fig:jrw_architecture}
\end{figure*}

\subsection{Sequential Walker Release}
When an EE freezes a node at time $t_0$, its $m$ trapped walkers are temporarily prevented from participating in the normal transport dynamics. These walkers enter a release queue and are returned to the mobile walker population sequentially. 

When $m\leq\Delta$, one walker is released at each of the final $m$ time steps of the frozen interval denoted as
\begin{equation}
(t_0+\Delta)-(m-1),\, ~ (t_0+\Delta)-(m-2),\, ~ \ldots,\, ~ t_0+\Delta.
\label{eq:release_schedule_small_m}
\end{equation}

For example, take $m=5$ and $\Delta=10$. The five walkers are released at $t_0+6,~t_0+7,~t_0+8,~t_0+9,~t_0+10$, so that exactly one walker is released at each of the final five time steps of the frozen interval $\Delta$.

When $m>\Delta$, the walkers are distributed as uniformly as possible over available release times,
\begin{equation}
m=r\Delta+s,
\label{eq:release_division}
\end{equation}
where $r$ is the number of walkers that can be released to the main stream of random walkers at each time step, and $s$ is the remaining number of walkers. 

For example, take $m=13$ and $\Delta=5$. We can then write $13=2(5)+3$, resulting in $r=2,~s=3$. The first $s=3$ time steps get $r+1=3$ walkers each, and the remaining $\Delta-s=5-3=2$ time steps get $r=2$ walkers each to be released to the main stream. Thus, the  distribution of the walkers in the $\Delta=5$ time steps will be $(3,3,3,2,2)$, with $3+3+3+2+2=13=m$.

Thus, the freezing mechanism does not release all trapped walkers simultaneously, instead the frozen walkers are gradually returned to the main stream of random walkers, providing a finite and controlled recovery of transport activity. This sequential release prevents the sudden re-injection of a large number of walkers into the network and thereby avoids artificially generating repeated EEs immediately following node recovery. For instance, a traffic jam at a junction (node) blocks both the junction and the vehicles (walkers). Each vehicle is released from  congestion at different times. 

\subsection{Degree-Resolved EE Probability}
Following Kishore \textit{et al.}~\cite{kishore2011}, we characterize the occurrence of EE by the degree-resolved probability $F(K)$, given in Eq.~\eqref{eq:FK_unbiased}. For RRW, we extend this quantity to
\begin{equation}
F_{\Delta}(K) = P\!\left[f_i(t)>q_i\,\middle| \tau_i(t)=0\right],
\label{eq:FDelta}
\end{equation}
where $\tau_i(t)=0$ denotes that the node $i$ is unfrozen at time $t$.
Thus, $F_{\Delta}(K)$ is the probability that an unfrozen node of
degree $K$ experiences a new EE. Frozen nodes are excluded from the conditioning ensemble, so repeated threshold exceedances do not contribute. For $\Delta=0$, all nodes remain unfrozen and Eq.~\eqref{eq:FDelta} reduces to the standard EE probability given in Eq.~\eqref{eq:FK_unbiased}.

\subsection{EE Frequency and Frozen Fraction}
The instantaneous frequency of new EEs is
\begin{equation}
F_{\mathrm{EE}}(t) = \sum_{i=1}^{N}\mathrm{EE}_i(t),
\label{eq:FEE}
\end{equation}
where $\mathrm{EE}_i(t)$ is given by Eq.~\eqref{eq:new_ee_condition}. Since an EE is counted only when it occurs at an active node, each counted EE corresponds to an active node entering the frozen state.

We characterize the macroscopic state of the jammed network (note that we use jammed and frozen interchangeably; both convey the same meaning) by the fraction of frozen nodes,
\begin{equation}
\phi(t) = \frac{N_{\mathrm{frozen}}(t)}{N},
\label{eq:frozen_fraction}
\end{equation}
where $N_{\mathrm{frozen}}(t)$ denotes the number of frozen nodes at time $t$. Therefore, the quantity $1-\phi(t)$ is the fraction of nodes that are unfrozen. While $F_{\mathrm{EE}}(t)$ quantifies the instantaneous generation of new EEs, $\phi(t)$ characterizes the population of  frozen nodes. 

\par
A visual representation of the RRW dynamics is shown in Fig.~\ref{fig:jrw_architecture}. In panel \ref{fig:jrw_architecture}(a), the walkers undergo conventional random-walk dynamics on the network. In panel \ref{fig:jrw_architecture}(b), an extreme event occurs when the number of walkers at a node exceeds its prescribed threshold, $f_i>q_i$. The affected node is subsequently frozen, immobilizing the walkers present there, as illustrated in panel \ref{fig:jrw_architecture}(c). The frozen node remains inaccessible for the prescribed duration $\Delta$.  The trapped walkers are released sequentially to neighboring nodes as the frozen interval ends, as shown in panel \ref{fig:jrw_architecture}(d). At $t=\Delta+1$, all the walkers have been released sequentially as illustrated in panel \ref{fig:jrw_architecture}(e). For this, $12$ walkers were simulated on a small BA network~\cite{barabasi1999scaling} with $N=10$ and $m=4$, with frozen interval $\Delta=10$.

\section{Analytical Framework}
\label{sec:Analytic}
We now develop a coarse-grained description of the feedback between EEs and frozen fraction $\phi(t)$. The latter modifies the effective population of mobile walkers, which in turn changes the EE probability and averaging it over the network gives the freezing rate, from which a delayed evolution equation for $\phi(t)$ follows. We subsequently introduce a linearized form of this rate to obtain analytically tractable transient and stationary results. All intermediate derivations are given in the Supplementary Material (see Notes~2--5), along with a short introduction to EE framework defined in Ref.~\cite{kishore2011} (see Note~1).

\subsection{Effective Mobile Walker Population}
Freezing a fraction $\phi(t)$ of the network does not remove the same fraction of walkers, because the stationary occupation of an unbiased random walk is degree-dependent. We therefore introduce the degree-bias factor (see Note~2(A) in Supplementary Material):
\begin{equation}
\kappa = \frac{\langle K\rangle_{\mathrm{freeze}}} {\langle K\rangle},
\label{eq:kappa_FP}
\end{equation}
where $\langle K\rangle_{\mathrm{freeze}}$ and $\langle K\rangle$ denote the mean degrees of frozen nodes and  that of the network, respectively. We then define the corresponding effective population of mobile walkers as (see Note~2(A) in Supplementary Material):
\begin{equation}
W_{\mathrm{eff}}(t) = W_0 \left[1-\kappa\phi(t)\right].
\label{eq:Weff}
\end{equation}

\subsection{Extreme-Event Probability with frozen nodes}
The occupation statistics evolve with $W_{\mathrm{eff}}$ while the threshold $q_i$ stays fixed at its value determined from $W_0$. The corresponding standardized threshold (see Note~2(B) in the Supplementary Material) measures how far the EE threshold lies above the mean node occupancy, in units of the standard deviation of the occupancy distribution:
\begin{equation}
z_i(\phi)  = \frac{q_i-W_{\text{eff}}p_i}{{\sqrt{W_{\text{eff}}p_i(1-p_i)}}},
\label{eq:zi_phi} 
\end{equation} 
where $p_i$ and $W_{\text{eff}}$ are taken from Eq.~\eqref{eq:stationary} and Eq.~\eqref{eq:Weff}, respectively.

Simplifying this further, we get (see Note~2(B) in the Supplementary Material):
\begin{equation}
z_i(\phi)=
\frac{M}{\sqrt{1-\kappa\phi}}
+\frac{\kappa\phi}{\sqrt{1-\kappa\phi}}
\sqrt{\frac{W_0p_i}{1-p_i}}.
\label{eq:sup_z_exact}
\end{equation}

Hence, under the Gaussian approximation, the EE probability of an active node $i$ is 
\begin{equation} 
\pi_i(\phi) = 1-\Phi\!\left[z_i(\phi)\right],
\label{eq:pi_phi} 
\end{equation} 
where $\Phi$ is the cumulative distribution function of the standard normal distribution (see Note~2(B) in Supplementary Material). At $\phi=0$, we have $W_{\mathrm{eff}}=W_0$ and Eq.~\eqref{eq:zi_phi} gives $z_i(0)=M$, recovering the original EE probability.

\subsection{Network-Averaged Freezing Rate}
The node-level EE probability in Eq.~\eqref{eq:pi_phi} can be averaged over the network to obtain
\begin{equation}
\overline{\pi}(\phi) = \frac{1}{N}\sum_{i=1}^{N}\pi_i(\phi).
\label{eq:pi_bar}
\end{equation}

Since the fraction of active nodes is $1-\phi$, the network-level rate of new freezing events is (see Note~2(C) in Supplementary Material):
\begin{equation}
R(\phi) = (1-\phi)\overline{\pi}(\phi).
\label{eq:R_phi}
\end{equation}

Note that in the absence of freezing, $\phi=0$, and the baseline freezing rate reduces to
\begin{equation}
R_0 = R(0) = \overline{\pi}(0).
\label{eq:R0}
\end{equation}

\subsection{Delayed Freezing Dynamics}
The network-level freezing rate $R(\phi)$ is directly related to the instantaneous frequency of newly occurring EEs. Since each counted EE freezes one previously active node, the number of nodes entering the frozen population at time $t$ is $F_{EE}(t)$, which is given by (see Note~3 in Supplementary Material):
\begin{equation}
F_{\mathrm{EE}}(t) = N R[\phi(t)].
\label{eq:FEE_R}
\end{equation}

Because each newly frozen node remains frozen for a duration $\Delta$, the frozen population at time $t$ consists of the nodes that have experienced a new EE during the preceding $\Delta$ time steps, which gives
\begin{equation}
\phi(t) = \frac{1}{N} \int_{t-\Delta}^{t} F_{\mathrm{EE}}(s)\,ds,
\label{eq:phi_FEE}
\end{equation}
with $F_{\mathrm{EE}}(s)=0$ at $t=0$. Differentiating Eq.~\eqref{eq:phi_FEE} gives
\begin{equation}
\frac{d\phi(t)}{dt} = \frac{1}{N}
\left[F_{\mathrm{EE}}(t) - F_{\mathrm{EE}}(t-\Delta)\right].
\label{eq:phi_dot_FEE}
\end{equation}
The first term represents the rate at which active nodes become frozen through newly occurring EEs, while the delayed term represents the rate at which previously frozen nodes become active again after their freezing interval has elapsed. 

Now, substituting $F_{\mathrm{EE}}(t)=NR[\phi(t)]$ yields the coarse-grained delayed evolution equation as
\begin{equation}
\frac{d\phi(t)}{dt} = R[\phi(t)] - \Theta(t-\Delta) R[\phi(t-\Delta)],
\label{eq:DDE_general}
\end{equation}
where $\Theta$ is the Heaviside step function. The delayed term accounts for the recovery of nodes whose freezing interval has elapsed, and for $t<\Delta$, it vanishes since no previously frozen node has yet completed its freezing interval. Equation~\eqref{eq:DDE_general} provides the macroscopic description of the RRW dynamics (see Note~3 in Supplementary Material).

\subsection{Linearized Freezing-Rate Approximation}
A complete Gaussian description of the probability  of EE is given by Eq.~\eqref{eq:pi_phi}, but its nonlinear dependence on $\phi$ makes the resulting network-averaged freezing rate difficult to study analytically. To obtain an analytically tractable form, we approximate its logarithmic variation about $\phi=0$. For node $i$, we get
\begin{equation}
\left. \frac{d\ln\pi_i}{d\phi} \right|_{\phi=0} = -\kappa c_i h(M),
\label{eq:log_slope}
\end{equation}
where
\begin{equation}
c_i = \frac{M}{2} + \sqrt{\frac{W_0p_i}{1-p_i}},
\end{equation}
and $h(M)$ is the inverse Mills ratio, denoting the corresponding Gaussian-tail sensitivity (see Note~4(A)--(B) in Supplementary Material).

Averaging \eqref{eq:log_slope} over the entire network results in
\begin{equation}
\overline{\pi}(\phi) \simeq R_0\exp\! \left[ -\kappa h(M)\langle c\rangle_w\phi \right],
\label{eq:pi_exp}
\end{equation}
where $R_0=\overline{\pi}(0)$ and $\langle c\rangle_w$ is the EE-weighted $c_i$ (see Note~4(C)--(D) in Supplementary Material).

The freezing rate can then be written as
\begin{equation}
R(\phi) \simeq (1-\phi)R_0 \exp[-\kappa h(M)\langle c\rangle_w\phi].
\label{eq:R_phi}
\end{equation}
Expanding $\ln(1-\phi)$ to first order in $\phi$ around $\phi=0$ yields
\begin{equation}
R(\phi) \simeq R_0 e^{-\beta\phi},
\label{eq:R_exp}
\end{equation}
with $\beta = 1+\kappa h(M)\langle c\rangle_w$ being an effective quantity inherited from the underlying degree-dependent EE statistics, not an independent fitting parameter. The reduction to the exponential form and the expression for $\beta$ are derived in the Supplementary Material (see Note~4(D)).

The approximation in Eq.~\eqref{eq:R_exp} assumes that the initial logarithmic sensitivity remains approximately constant over the range of $\phi$ explored by the dynamics; its accuracy is therefore expected to deteriorate at sufficiently large values of $\Delta$ and thus for large frozen fractions.

Substituting Eq.~\eqref{eq:R_exp} in Eq.~\eqref{eq:DDE_general}, we get the nonlinear delay differential equation (DDE) as
\begin{equation}
\frac{d\phi(t)}{dt} = R_0 e^{-\beta\phi(t)} - \Theta(t-\Delta) R_0 e^{ -\beta\phi(t-\Delta)}.
\label{eq:DDE_exponential}
\end{equation}
The usefulness of Eq.~\eqref{eq:DDE_exponential} is that it separate the two essential mechanisms of the dynamics: the slowdown of $\phi$ due to nonlinear braking through $e^{-\beta\phi}$ and delayed recovery through $\phi(t-\Delta)$. Although this equation is already an approximation to the underlying microscopic closure, it retains the nonlinear dependence on $\phi$ and remains analytically tractable while preserving the structure of the RRW dynamics. The subsequent method-of-steps analysis allows the effects of the delay to be isolated explicitly.

\begin{figure*}[t]
    \centering
    \includegraphics[width=\textwidth]{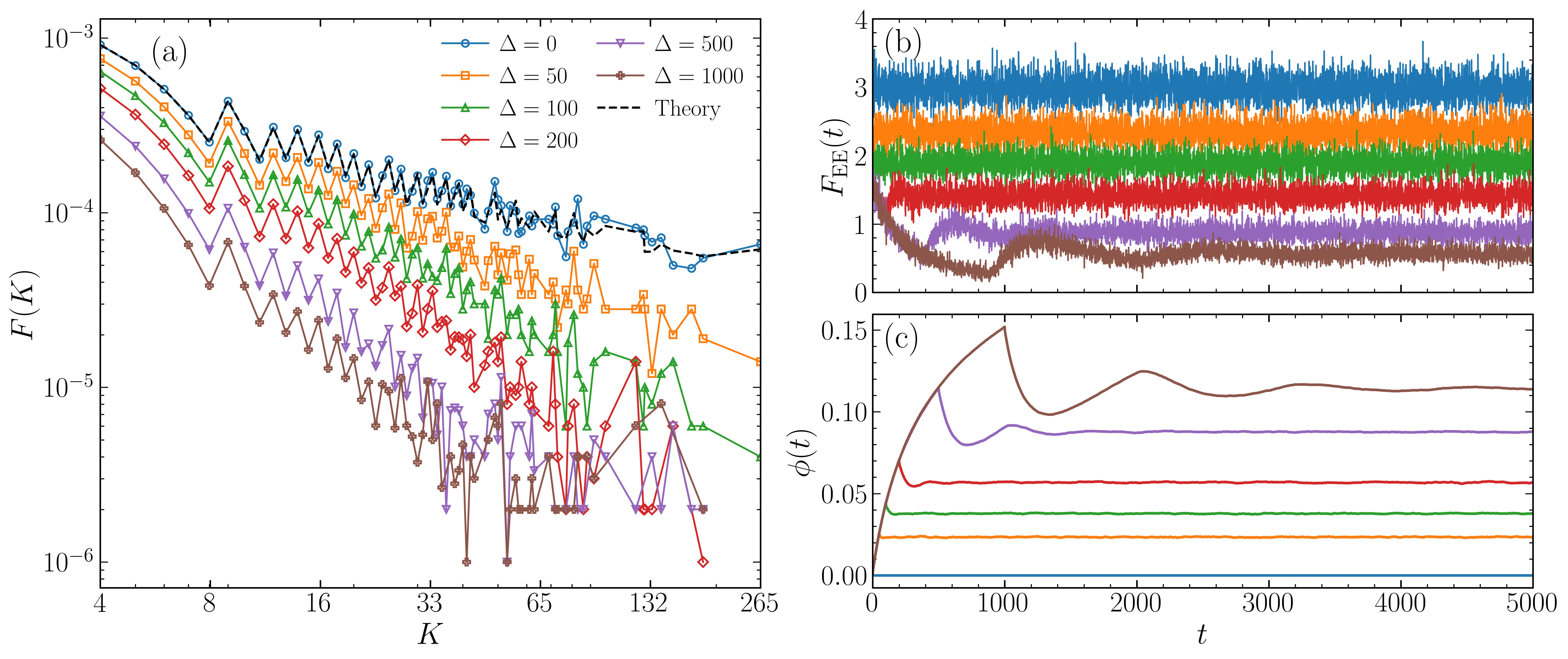}
    \caption{Effect of the freezing time $\Delta$ on the EE dynamics. (a) Degree-dependent EE probability $F(K)$. The dashed curve shows the theoretical prediction for the unbiased random walk, given in Eq.~\eqref{eq:FK_unbiased} and rest are the numerical results. (b) Instantaneous number of newly generated EEs, $F_{\mathrm{EE}}(t)$. (c) Temporal evolution of the frozen-node fraction $\phi(t)$. In all the panels, blue curve is for $\Delta=0$, orange curve for $\Delta=50$, green curve for $\Delta=100$, red curve for $\Delta=200$, purple curve for $\Delta=500$, and brown curve for $\Delta=1000$. All simulations were done on BA networks~\cite{barabasi1999scaling} with $N=5000, m=4$, for $T=5000$ time steps, while omitting $100$ initial transient steps, and averaged over $100$ realizations.}
    \label{fig:main_results}
\end{figure*}

\subsection{Analytical Quantities}
The initial history is taken to be
\begin{equation}
\phi(t)=0, \quad -\Delta\leq t\leq0.
\label{eq:initial_history}
\end{equation}

For the first interval $0\leq t\leq\Delta$, the delayed contribution is absent and the equation admits the closed-form solution (see Note~5(A) in Supplementary Material):
\begin{equation}
\phi(t) = \frac{1}{\beta} \ln\!\left(1+\beta R_0t\right).
\label{eq:first_interval}
\end{equation}
During the first interval, every node that freezes remains frozen throughout the interval, which is why the solution is independent of $\Delta$. The logarithmic form therefore describes the intrinsic growth of the frozen fraction under the nonlinear braking mechanism alone. At early times $(\beta R_0 t\ll1)$, it reduces to $\phi(t)\simeq R_0t$ elucidating that the initial growth is approximately linear with slope equal to the baseline freezing rate $R_0$. At later times within the same interval the logarithmic dependence causes the growth to slow down, thus new EEs are suppressed as $\phi(t)$ increases. The initial trajectory provides a direct measure of the freezing kinetics independently of the delay, which is why $R_0$ and $\beta$ can be determined solely from the pre-delay dynamics.

For the second interval $\Delta\leq t\leq2\Delta$, writing $\tau=t-\Delta$ (see Note~5(B) in Supplementary Material):
\begin{equation}
\phi(\Delta+\tau) =
\frac{1}{\beta} \ln\!\left[ \frac{1+\beta R_0\Delta+\beta R_0\tau+\frac{1}{2}(\beta R_0)^2\tau^2}
{1+\beta R_0\tau}\right],
\label{eq:second_interval}
\end{equation}
with $0\leq\tau\leq\Delta.$
The second-interval solution shows how the delayed recovery process modifies the initial growth. Unlike the first interval, the evolution now contains information about the freezing history through the delayed term $R(\tau)$. The numerator in Eq.~\eqref{eq:second_interval} contains the contribution accumulated during the interval since $t=0$, while the denominator accounts for the nodes that  begun to recover. Thus, the expression describes the competition between ongoing freezing and the release of nodes that have frozen earlier, which is the origin of the first peak and subsequent decrease of $\phi(t)$. The method of steps therefore shows that each successive interval incorporates another portion of the freezing history into the present dynamics.

The minimum following the first maximum occurs when the instantaneous freezing and delayed recovery rates balance, giving the trough time (see Note~5(C) in Supplementary Material):
\begin{equation}
t_{\rm trough} = \Delta + \frac{\sqrt{1+2\beta R_0\Delta}-1} {\beta R_0},
\label{eq:trough_time}
\end{equation}
with corresponding frozen fraction
\begin{equation}
\phi_{\rm trough} = \frac{1}{2\beta}\ln\!\left(1+2\beta R_0\Delta\right).
\label{eq:trough_phi}
\end{equation}
These expressions show that the first trough is determined entirely by the competition between the instantaneous freezing dynamics and the delayed recovery dynamics. The trough is reached when the two rates become equal: the number of newly frozen nodes per unit time is exactly balanced by the number of nodes being released from the frozen state. Before this point, the rate of recovery dominates and $\phi(t)$ decreases from its first maximum; after the balance is crossed the instantaneous freezing rate exceeds the delayed recovery rate again, initiating the next growth phase. Consequently, the analytical trough prediction provides a direct test of whether the reduced exponential closure captures the competition between these two timescales.

It is important to note that the linearized approximation and the corresponding method-of-steps solutions are introduced solely to obtain analytical insight into the delayed dynamics. Therefore, the numerical comparisons presented in Sec.~\ref{sec:Results} are performed using the full non-approximated first-principles expression.

\section{Results}
\label{sec:Results}

In this section, we validate the theoretical framework using numerical simulations of the proposed RRW model. We first examine how the frozen time $\Delta$ modifies the degree-resolved $F(K)$ and temporal extreme-event $F_{EE}$ statistics, then investigate their microscopic organization on the network, and finally compare the observed transient dynamics with the analytical predictions for the first two delay intervals. Unless stated otherwise, all simulations are performed on networks of size $N=5000$, with $m=4$ edges added per newly introduced node, for $T=5000$ time steps, while discarding $100$ initial transient steps, and averaged over $100$ independent network realizations.

\begin{figure*}[t]
    \centering
    \includegraphics[width=\textwidth]{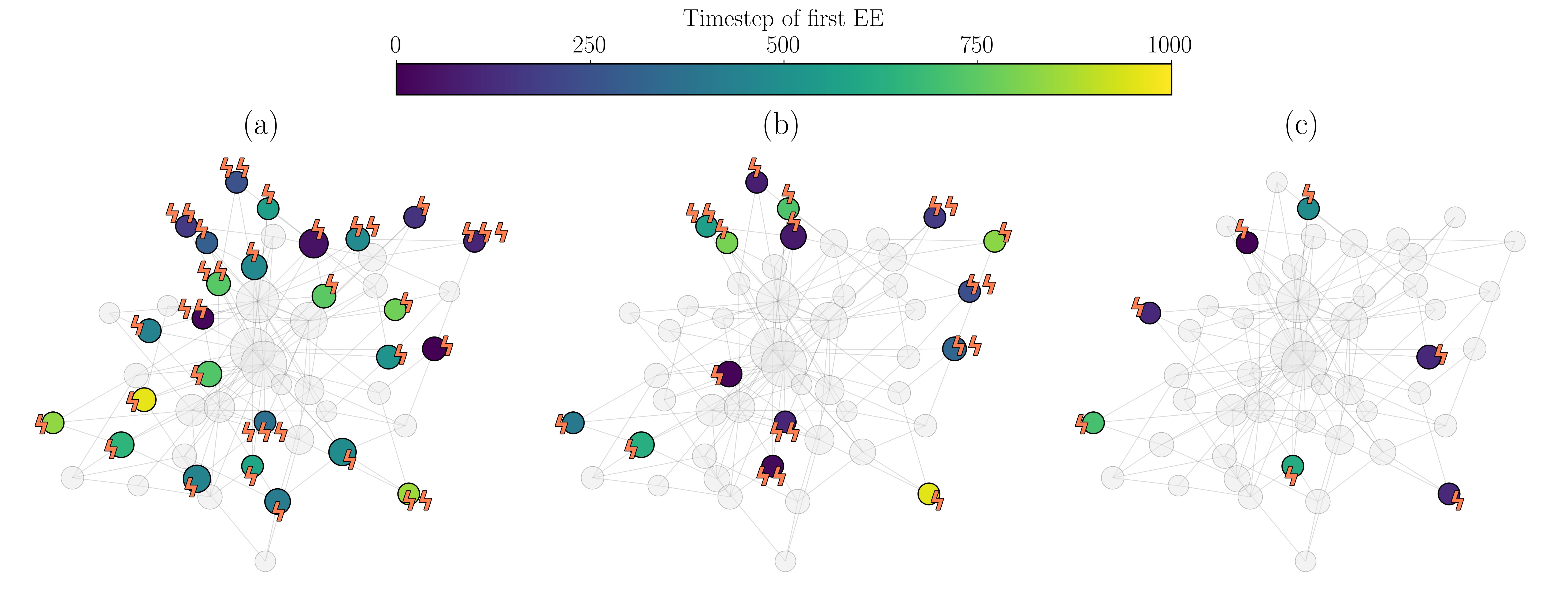}
    \caption{Microscopic dynamics of RRW for three freezing times, (a) $\Delta=0$, (b) $\Delta=100$, and (c) $\Delta=1000$, on the same representative BA network with $N=50$ and $m=4$. Node size is proportional to degree, the color of each affected node indicates the timestep at which its first EE occurs, and the number of lightning symbols (\textbf{\textcolor{orange}{Ϟ}}) is equal to the number of EEs that occured at that node. Nodes that have not experienced an EE are shown in gray.}
    \label{fig:jrw_genuine_EE}
\end{figure*}

\subsection{Effect of the frozen time on EE dynamics}
We first study how the introduction of the freezing mechanism affects the EE dynamics of the proposed recovery random walk (RRW) model. We unravel the effect on the degree-dependent EE probability $F(K)$, the instantaneous frequency of EEs $F_{EE}(t)$, and the fraction of frozen nodes $\phi(t)$, as shown in Fig.~\ref{fig:main_results}.

\subsubsection{Degree-dependent EE probability}
Figure~\ref{fig:main_results}(a) shows the probability $F(K)$ of a node with degree $K$ to experience an extreme event. For $\Delta=0$, the simulation reproduces the characteristic degree dependence of the simple random walk EE framework: $F(K)$ decreases with increasing degree, implying that low-degree nodes are more likely to experience EEs than hubs~\cite{kishore2011}. The black dashed curve in Fig.~\ref{fig:main_results}(a) is the theoretical result~\eqref{eq:FK_unbiased} for the simple random walk EE framework.

Introducing a frozen duration does not qualitatively change this degree-dependence. The curves for different $\Delta$ keep the same overall decreasing trend with $K$, while their magnitude progressively decreases as $\Delta$ increases. Thus, the primary effect of the frozen duration is a suppression of EE probability, not a change in its degree dependence. This suppression originates from the fact that EEs cause freezing, and thus increase the frozen node fraction $\phi(t)$, which traps more walkers and in turn reduces the effective walker population $W_{\text{eff}}$ (see Eq.~\eqref{eq:Weff}). This effect reduces the probability of EE at active nodes due to lack of enough mobile walkers to cause more EEs, thus resulting in a negative feedback loop. The degree-dependent EE probability curve inherited from the underlying random walk therefore persists in the recovery model, while the overall magnitude is reduced as $\Delta$ increases.

\subsubsection{EE frequency}
The instantaneous number of newly generated EEs at time $t$, $F_{EE}(t)$, is depicted in Figure~\ref{fig:main_results}(b). Note that repeated EEs at the frozen node are not counted, and the plotted quantity measures EEs occurring only at the active nodes. Since every counted EE freezes an active node, $F_{EE}(t)$ directly measures the rate at which nodes enter the frozen population.
It should also be noted that the same color legends are used for different frozen duration $\Delta$.

For $\Delta=0$, the EE frequency fluctuates around a stationary level, with no feedback since there are no frozen nodes. As $\Delta$ increases, the stationary EE frequency systematically decreases, consistent with the suppression of $F(K)$ observed in Fig.~\ref{fig:main_results}(a). The decrease in $F_{EE}(t)$ is a result of the same negative feedback mechanism stated earlier: EEs increase $\phi(t)$, which reduces $W_{\text{eff}}$ and consequently suppresses further EE generation. For large $\Delta$,  a pronounced transient behavior appears: EE frequency initially decreases, reaching a minimum around $t=\Delta$, then recovers and undergoes damped oscillations before approaching a stationary level. These oscillations arise from the delayed influence of previously frozen nodes, as they remain frozen for $\Delta$ time steps before thawing and becoming active.

\begin{figure*}[t]
\centering
\includegraphics[width=0.75\textwidth]{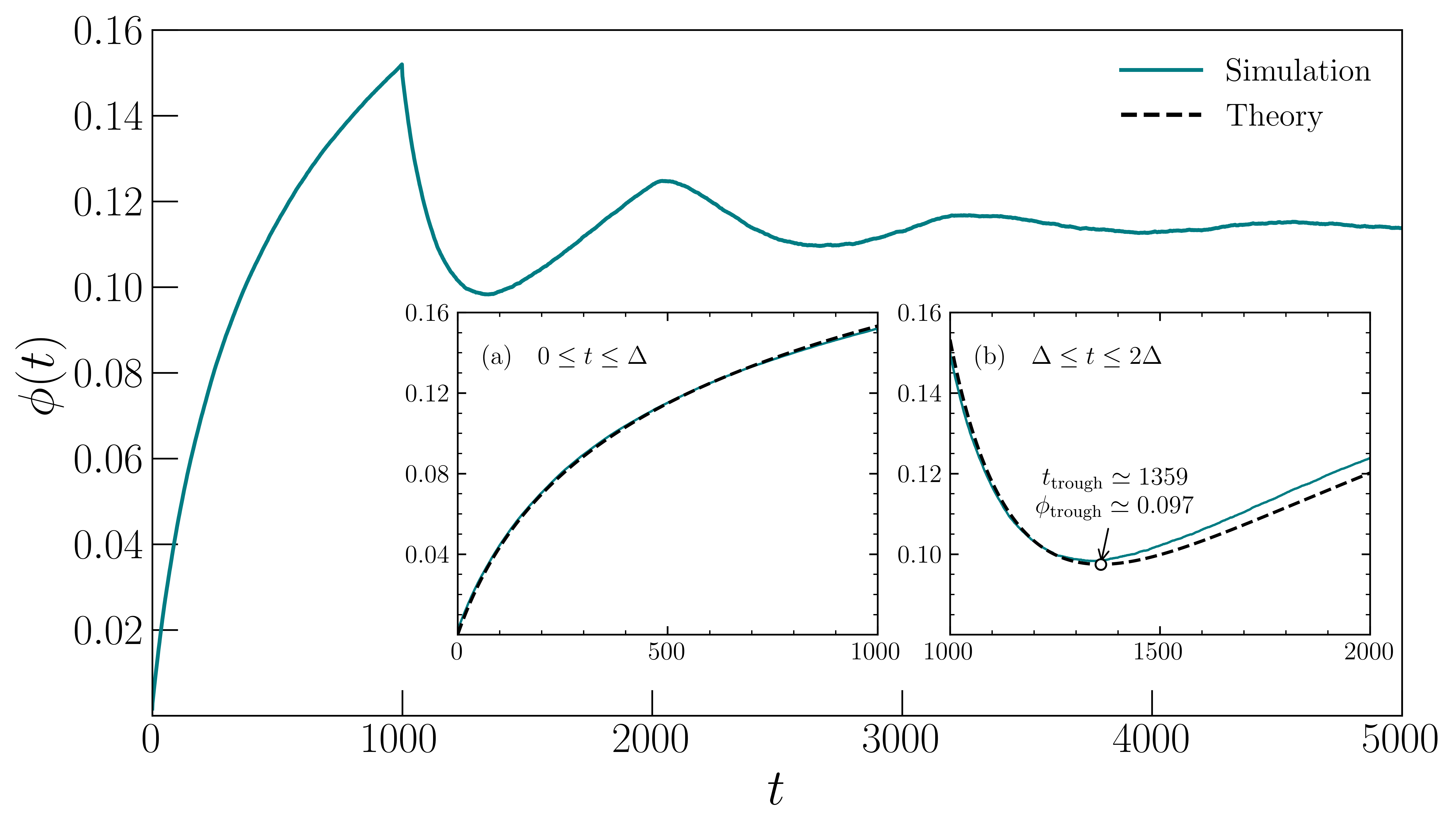}
\caption{Analytical description of the transient frozen node fraction for $\Delta=1000$. The solid teal curve shows the simulation result, while the black dashed curves in the insets show the non-approximated analytical prediction. Inset (a) shows the first interval, $0\leq t\leq\Delta$, where the dynamics are governed solely by the instantaneous freezing rate. Inset (b) shows the second interval, $\Delta\leq t\leq2\Delta$, where the delayed unfreezing term becomes active. The analytical trough point is highlighted in inset (b), with its time and frozen-node fraction values indicated.}
\label{fig:method_steps}
\end{figure*}

\subsubsection{Frozen node fraction}
Figure~\ref{fig:main_results}(c) illustrates the temporal evolution of the fraction of the frozen node $\phi(t)$. For $\Delta=0$, $\phi(t)=0~\forall~t$, since no nodes are frozen. For finite $\Delta$, $\phi(t)$ initially increases as new EEs progressively freeze active nodes, which then decreases after $\Delta$ time due to release of the previously frozen nodes,
and then gradually saturates, which we denote by $\phi^*$. This stationary value increases with $\Delta$, since a larger freezing duration allows frozen nodes to stay frozen for a longer time.

For small $\Delta$, $\phi(t)$ rapidly approaches its stationary value with a weak transient, but as $\Delta$ increases the transient dynamics is more pronounced. In particular, for $\Delta=500~\&~1000$, $\phi(t)$ reaches a maximum around the first freezing time, then decreases as the frozen nodes during the earlier cycle begin to thaw. Subsequent recovery and damped oscillations arise from the competition between instantaneous freezing and delayed unfreezing rates. Eq.~\eqref{eq:phi_dot_FEE} from the analytical description shows that $\phi(t)$ increases when the current EE generation rate (or the node freezing rate) exceeds the rate at which previously frozen nodes are released, decreases when the delayed rate dominates, and reaches an extremum when the two rates become equal. Because the unfreezing term contains the delayed time $t-\Delta$, this competition generates a characteristic oscillation timescale of order $\Delta$. Thus, successive extrema are separated by approximately $\Delta$, while their amplitude gradually decreases as the system approaches the stationary state.

\subsection{Microscopic organization of EEs}
Note that  $F_{EE}(t)$ provides the total frequency of newly occurring EEs, however, its microscopic origin is associated with the onset of EE at a specific node and at a particular time. To examine this microscopic organization, we take a representative BA network~\cite{barabasi1999scaling} with $N=50$ and $m=4$, and we record the time at which each node experiences its first EE. The resulting networks are shown in Fig.~\ref{fig:jrw_genuine_EE} for $\Delta=0,100~\&~1000$, at the end of the observation window $t=1000$. The node size is proportional to its node degree, while the color indicates the time of its first EE; nodes that do not experience EEs during the observation window are shown in gray. The number of lightning symbols (\textbf{\textcolor{orange}{Ϟ}}) at a node is equal to the number of EEs experienced by that node within the observation window $t=1000$. Recall that the repeated EEs in the frozen nodes are not counted.

For the case $\Delta=0$ (see Fig.~\ref{fig:jrw_genuine_EE}(a)), a large fraction of the network experiences EEs within the observation window. Increasing $\Delta$ progressively suppresses the occurrence of EEs at individual nodes. For $\Delta=100$ (see Fig.~\ref{fig:jrw_genuine_EE}(b)), the fraction of unaffected nodes increases, and this effect becomes more pronounced for $\Delta=1000$ (see Fig.~\ref{fig:jrw_genuine_EE}(c)), where only a small subset of nodes experience EE within the observation window. Thus, increasing the frozen duration does not merely alter the timing of EE activity, but rather reduces the set of nodes that can experience an EE. This microscopic suppression is, again, a direct consequence of the reduction in effective mobile walker population: increasing $\Delta$ prolongs walker immobilization, and consequently reduces the likelihood of subsequent EEs at active nodes.

These results provide insights on the microscopic counterpart of the reduction in $F_{EE}(t)$ observed in Fig.~\ref{fig:main_results}(b). While $F_{EE}$ measures the total number of newly occurring EEs at each time, the graphs in Figs.~\ref{fig:jrw_genuine_EE}(a)-\ref{fig:jrw_genuine_EE}(c) resolve how these events are distributed among individual nodes and reveal the progressive reduction in number of nodes participating in EE activity as $\Delta$ increases.

\subsection{Analytical description of the frozen node population}
Now we compare the analytical framework with the simulated dynamics of the fraction of the frozen nodes. We focus on $\Delta=1000$, where the transient and damped relaxation is more pronounced. The comparison uses the non-approximated first-principles prediction; the linearized method-of-steps solution derived in Sec.~\ref{sec:Analytic} is used only for mathematical interpretation and is not used to generate the analytical curves shown in Fig.~\ref{fig:method_steps}.

For the first interval $0\leq t\leq\Delta$, no previously frozen node has yet completed its frozen interval and the dynamics are governed solely by the instantaneous freezing rate. The resulting theoretical solution agrees closely with the simulation throughout this interval, as shown in inset (a) of Fig.~\ref{fig:method_steps}. The first maximum occurs at $t=\Delta$, with $\phi_{\mathrm{peak}}^{\mathrm{sim}}\simeq0.152$ and $\phi_{\mathrm{peak}}^{\mathrm{th}}\simeq0.153$. Thus, the analytical description accurately captures the initial accumulation of frozen nodes, and the location and magnitude of the first maximum.

For the second interval $\Delta\leq t\leq2\Delta$, the delayed unfreezing term in \eqref{eq:DDE_general} becomes active. The theoretical prediction again closely follows the simulated trajectory, reproducing both the decrease in $\phi$ and its subsequent recovery, as shown in inset (b). The simulation reaches a minimum or trough, $\phi_{\mathrm{trough}}^{\mathrm{sim}}\simeq0.098$ at $t_{\mathrm{trough}}^{\mathrm{sim}}\simeq1345$, while the analytical prediction gives $\phi_{\mathrm{trough}}^{\mathrm{th}}\simeq0.097$ at $t_{\mathrm{trough}}^{\mathrm{th}}\simeq1359$. 
Once again, theoretically predicted characteristic timescale response as well as its amplitude are in good agreement with the simulation.
\\

Beyond the second interval, the temporal evolution of the fraction of the frozen nodes exhibits further damped oscillations, which result from the competing interaction between freezing and unfreezing dynamics,  before approaching the stationary state
and are governed by the delay differential equation~\eqref{eq:DDE_general}. The close agreement over the first two delay intervals therefore demonstrates that the coarse-grained analytical framework captures both the initial accumulation of frozen nodes and the first delayed relaxation induced by the finite freezing duration. The same procedure can be iterated over successive delay intervals using Eq.~\eqref{eq:DDE_general}, yielding the theoretical prediction for $\phi(t)$ at arbitrary times matching exactly with that shown numerically in Fig.~\ref{fig:main_results}(c). Together, these interval-wise solutions constitute the complete analytical description of the fraction of the frozen nodes.

\section{Discussion and Conclusion}
\label{sec:DiscConc}

The proposed recovery random walk (RRW) extends the conventional simple random walk by allowing local accumulation due to EE and  temporally modify the transport dynamics. Following an EE, the affected node is frozen for a finite duration, temporarily immobilizing the walkers present there. Physically, this provides a realistic description of networked transport systems in which sufficiently large local loads can induce temporary loss of mobility or accessibility. The framework is therefore applicable to generic transport processes involving particles, agents, vehicles, or information.

The degree dependence of EEs observed for conventional random walks is retained in the RRW, with low-degree nodes remaining more likely to experience extreme events than highly connected nodes. As shown in Fig.~\ref{fig:main_results}(a), increasing the freezing duration systematically reduces the magnitude of the degree-dependent EE probability while preserving its overall 
degree dependent dynamics. The total EE frequency, in turn, is suppressed as $\Delta$ increases (Fig.~\ref{fig:main_results}(b)), while the stationary frozen-node fraction increases (Fig.~\ref{fig:main_results}(c)). Thus, increasing the freezing duration primarily changes the magnitude and frequency of EE activity rather than its underlying degree dependent dynamics.

The finite frozen duration introduces a negative feedback between EE generation and subsequent transport activity. EEs remove walkers from the mobile population, reducing the likelihood of further events, while the eventual release of previously frozen nodes restores mobility. This delayed recovery produces a pronounced transient response for sufficiently large $\Delta$, with an initial accumulation of frozen nodes followed by relaxation and damped oscillations. The analytical framework captures these features closely, as shown in Fig.~\ref{fig:method_steps}.
These results demonstrate that the delayed coarse-grained description captures both the characteristic timescale and amplitude of the transient response.

Overall, the RRW provides a realistic framework for studying extreme events in transport processes where local extremes can temporarily alter network accessibility. The results show that freezing suppresses subsequent EE activity while preserving the degree dependence inherited from the underlying unbiased random walk, and that finite recovery times introduce delayed feedback and transient oscillatory dynamics. The close agreement between simulations and the first-principles analytical description demonstrates that these effects can be captured within a tractable coarse-grained framework.

The present study considers unbiased random walks on BA networks with deterministic freezing duration. Future work could therefore investigate heterogeneous or degree-dependent freezing times, alternative network topologies, and biased random walks, where the interplay between transport bias and freezing may produce qualitatively different extreme-event statistics. Extending the framework to more general waiting-time distributions could also provide a natural extension toward describing systems with heterogeneous relaxation times.

\section{Supplementary Material}
The supplementary material provides detailed mathematical derivations supporting the analytical framework developed in the main text. It includes the baseline extreme event statistics for unbiased random walks, the derivation of the effective mobile walker population in the presence of frozen nodes, the derivation of the full Gaussian extreme event probability, and the network-averaged freezing rate. It further presents the derivation of the delayed freezing dynamics, the linearization of the network-averaged freezing rate, and the analytics describing the transient dynamics of the reduced delay differential equation.

\section{Acknowledgements}
K.S. and N.R.V. acknowledge financial support from IISER-TVM. V.K.C. is supported by the ANRF project under Grant No. ANRF/ARG/2025/004108/PS. D.V.S is supported by the ANRF Project under Grant No. ANRF/ARG/2025/008542/PS. 
We thank Dipranjan Pal for the insightful initial discussions about the work.

\section{Data Availability}
The data supporting the findings of this article are not publicly available. The data that support the findings of this study are available from the corresponding author upon reasonable request.

\nocite{*}
\bibliography{references}

\end{document}